\documentclass[runningheads]{llncs}
\usepackage{graphicx}
\usepackage{graphicx}
\usepackage{amsmath}
\usepackage{booktabs}
\usepackage{algorithm}
\usepackage{algorithmic}
\usepackage{amsfonts}
\usepackage{multirow}
\usepackage{makecell}
\usepackage{subfigure}
\usepackage{color}
\usepackage{bm}
\usepackage{epstopdf}
\usepackage{url}
\usepackage{balance}
\usepackage{threeparttable}
\usepackage[misc]{ifsym}

\begin{document}
\title{Deep Structural Point Process for Learning Temporal Interaction Networks}
\titlerunning{DSPP for Learning Temporal Interaction Networks}
%
%

\author{Jiangxia Cao\inst{1,2} \and
Xixun Lin\inst{1,2} (\Letter)\and
Xin Cong\inst{1,2} \and
Shu Guo\inst{3} \and
Hengzhu Tang\inst{1,2} \and
Tingwen Liu\inst{1,2} (\Letter)\and
Bin Wang\inst{4}
}

\authorrunning{Cao et al.}
\tocauthor{Jiangxia Cao, Xixun Lin, Xin Cong, Shu Guo, Hengzhu Tang, Tingwen Liu, Bin Wang}
\toctitle{Deep Structural Point Process for Learning Temporal Interaction Networks}

\institute{Institute of Information Engineering, Chinese Academy of Sciences \and School of Cyber Security, University of Chinese Academy of Sciences
\and National Computer Network Emergency Response Technical Team/Coordination Center of China
\and Xiaomi AI Lab, Xiaomi Inc. \\
\email{\{caojiangxia,linxixun,congxin,tanghengzhu,liutingwen\}@iie.ac.cn},
\email{guoshu@cert.org.cn},
\email{wangbin11@xiaomi.com}}

\maketitle              

\begin{abstract}
This work investigates the problem of learning temporal interaction networks. A temporal interaction network consists of a series of chronological interactions between users and items. Previous methods tackle this problem by using different variants of recurrent neural networks to model interaction sequences, which fail to consider the structural information of temporal interaction networks and inevitably lead to sub-optimal results. To this end, we propose a novel \textbf{D}eep \textbf{S}tructural \textbf{P}oint \textbf{P}rocess termed as \textbf{DSPP} for learning temporal interaction networks. DSPP simultaneously incorporates the \emph{topological structure} and \emph{long-range dependency structure} into the intensity function to enhance model expressiveness. To be specific, by using the topological structure as a strong prior, we first design a topological fusion encoder to obtain node embeddings. An attentive shift encoder is then developed to learn the long-range dependency structure between users and items in continuous time. The proposed two modules enable our model to capture the user-item correlation and dynamic influence in temporal interaction networks. DSPP is evaluated on three real-world datasets for both tasks of item prediction and time prediction. Extensive experiments demonstrate that our model achieves consistent and significant improvements over state-of-the-art baselines.
\end{abstract}

\section{Introduction}
Temporal interaction networks are useful resources to reflect the relationships between users and items over time, which have been successfully applied in many real-world domains such as electronic commerce~\cite{bigi}, online education~\cite{liyanagunawardena2013moocs} and social media~\cite{iba2010analyzing}. A temporal interaction network naturally keeps a graph data structure with temporal characteristics, where each edge represents a user-item interaction marked with a concrete timestamp. 


%
Representation learning on temporal interaction networks has gradually become a hot topic in the research of machine learning~\cite{survey}. A key challenge of modeling temporal interaction networks is how to capture the evolution of user interests and item features effectively. Because users may interact with various items sequentially and their interests may shift in a period of time. Similarly, item features are also ever-changing and largely influenced by user behaviours. Recent works have been proposed to tackle this challenge by generating the dynamic embeddings of users and items~\cite{ctdne,coeve,jodie,dgcf}. Although these methods achieve promising results to some extent, they still suffer from the following two significant problems.
1) \textbf{Topological structure missing}.
Most previous methods regard learning temporal interaction networks as a coarse-grained sequential prediction problem and ignore the topological structure information.
%
In fact, instead of only treating a temporal interaction network as multiple interaction sequences, we can discover user similarity and item similarity from the view of the topological structure. Nevertheless, due to the bipartite nature of temporal interaction networks, each node is not the same type as its adjacent nodes, so that we have to develop a flexible method to capture such a meaningful topology.
%
%
2) \textbf{Long-range dependency structure missing}. Most current methods are built upon the variants of recurrent neural networks (RNNs) to learn interaction sequences. Hence, they typically pay more attention to short-term effects and miss the dependency structure in long-range historical information~\cite{attention,bert}. But learning the long-range dependency structure in temporal interaction networks is also critical, since it can better model the long-standing user preference and intrinsic item properties.
In this paper, we propose the \textbf{D}eep \textbf{S}tructural \textbf{P}oint \textbf{P}rocess termed as \textbf{DSPP} to solve above problems.
%
Following the framework of Temporal Point Process (TPP)~\cite{ppsurvey}, we devise a novel intensity function which combines the topological structure and the long-range dependency structure to capture the dynamic influence between users and items.
%
%
%
%
%
Specifically, we first design a topological fusion encoder (TFE) to learn the topological structure. TFE includes a two-steps layer to encourage each node to aggregate homogeneous node features.
To overcome the long-range dependency issue, we then develop an attentive shift encoder (ASE) to recognize the complex dependency between each historical interaction and the new-coming interaction. 
Finally, we incorporate the learned embeddings from TFE and ASE into our intensity function to make time prediction and item prediction.
The main contributions of our work are summarized as follows: 
\begin{itemize}
\item We propose the novel DSPP to learn temporal interaction networks within the TPP paradigm, with the goal of solving two above structural missing problems simultaneously.
\item DSPP includes the well-designed TFE and ASE modules. TFE utilizes the topological structure to generate steady embeddings, and ASE exploits the long-range dependency structure to learn dynamic embeddings. Furthermore, these two types of embeddings can be seamlessly incorporated into our intensity function to achieve future prediction.
\item Extensive experiments are conducted on three public standard datasets. Empirical results show that the proposed method achieves consistent and significant improvements over state-of-the-art baselines\footnote{The source code is available from \url{https://github.com/cjx96/DSPP}.}. 
\end{itemize}

\section{Related Work}
\label{related}
Previous studies for learning temporal interaction networks can be roughly divided into the following three branches: random walk based method, RNN based methods and TPP based method. 
\begin{itemize}

\item Random walk based method. Nguyen $et$ $al$. propose CTDNE~\cite{ctdne} which models user and item dynamic embeddings by the random walk heuristics with temporal constraints. CTDNE first samples some time increasing interaction sequences, and then learns context node embeddings via the skip-gram algorithm~\cite{word2vec}. However, it ignores the useful time information of the sampled sequences, e.g., a user clicks an item frequently may indicate that the user pays more attention to this item at the moment.

\item RNN based methods. LSTM~\cite{lstm}, RRN~\cite{rrn}, LatentCross~\cite{latent} and Time-LSTM~\cite{timelstm} are pioneering works in this branch. For example, RRN provides a unified framework for combining the static matrix factorization features with the dynamic embeddings based on LSTM. Moreover, it provides a behavioral trajectory layer to project user and item embeddings over time. LatentCross is an extension of the architecture of GRU~\cite{gru}, which incorporates multiple types of context information. Time-LSTM develops the time gates for LSTM for modeling the interaction time information. Furthermore, JODIE~\cite{jodie} and DGCF~\cite{dgcf} are the state-of-the-art methods for learning temporal interaction networks via the coupled variants of RNNs. JODIE defines two-steps embedding update operation and an embedding projection function to predict the target item embedding for each user directly. DGCF extends JODIE by considering the 1-hop neighboring information of temporal interaction networks. 
\item TPP based method. DeepCoevolve~\cite{coeve} is a promising work that applies TPPs to learn temporal interaction networks. It uses a multi-dimensional intensity function to capture the dynamic influence between users and items. However, DeepCoevolve maintains the same embeddings of user and item until it involves a new interaction, which is not consistent with real-world facts~\cite{jodie}. Furthermore, it uses a linear intensity function to describe the dynamic influence, leading to the limited model expressiveness. 
\end{itemize}

\section{Background}
\subsection{Temporal Interaction Network}
\label{TIN}
A series chronological interactions can be represented as a temporal interaction network. Formally, a temporal interaction network on the time window $[0,T)$ can be described as $\mathcal{G}(T)=(\mathcal{U},\mathcal{V},\mathcal{E})$, where $\mathcal{U}$, $\mathcal{V}$ and $\mathcal{E}$ denote the user set, item set and interaction set, respectively. Each element $(u_i, v_j, t) \in \mathcal{E}$ is an interaction, describing that the user $u_i \in \mathcal{U}$ conducts an action with the item $v_j \in \mathcal{V}$ at the concrete timestamp $t$.

\subsection{Temporal Point Process}
\label{TPP}
TPPs are one of branches of stochastic processes for modeling the observed random discrete events, e.g. user-item interactions over time. 
Using conditional intensity function $\lambda(t)$ is a convenient way to describe TPPs. Given a interaction sequence that only has time information $\mathcal{T}:=\{t_i\}_{i=1}^{n}$ and an infinitesimal time interval $\mathrm{d}t$, where $t_i \in \mathbb{R}^+$ and $0 < t_1 < t_2 ... < t_n$, $\lambda(t)\mathrm{d}t$ is the conditional probability of happening an interaction in the infinitesimal interval $[t,t+\mathrm{d}t)$ based on $\mathcal{T}$. It can be also interpreted heuristically in the following way:
\begin{equation*}
\small
\begin{split}
\lambda(t)\mathrm{d}t := \mathbb{P}\{\text{an interaction occurs in } [t, t+\mathrm{d}t)|\mathcal{T}\} = \mathbb{E}[N([t,t+\mathrm{d}t))|\mathcal{T}],
\end{split}
\label{lambda}
\end{equation*}
where $N([t,t+\mathrm{d}t))$ is used to count the number of interactions happened in the time interval $[t,t+\mathrm{d}t)$. A general assumption made here is that there is either zero or one interaction happened in this infinitesimal interval, i.e., $N([t,t+\mathrm{d}t)) \in \{0,1\}$. Furthermore, given a future timestamp $t^+ > t_n$ and $\mathcal{T}$, we can formulate the conditional probability of no interactions happened during $[t_n, t^+)$ as $S(t^+)=\exp{(-\int_{t_n}^{t^+}\lambda(t)\text{d}t)}$. Therefore, the conditional probability density of the next interaction happened at the future timestamp $t^+$ can be defined as: $f(t^+) = S(t^+)\lambda(t^+)$, which means that no interaction happens in time interval $[t_n,t^+)$ and an interaction happens in infinitesimal interval $[t^+, t^+ + \mathrm{d}t)$. 



Multivariate Hawkes Process (MHP) is one of the most important TPPs for modeling interaction sequences~\cite{hawkes}. We denote $\mathcal{S}_{u_i}(T) = (\mathcal{V},\mathcal{H}_{u_i})$ as an interaction sequence of user $u_i$ on the time window $[0,T]$, where $\mathcal{H}_{u_i}$ is the interaction sequence of user $u_i$. The $h$-th interaction of $\mathcal{H}_{u_i}$ is denoted as $(v^h, t^h)$, which describes that the user $u_i \in \mathcal{U}$ has interacted with the item $v^h \in \mathcal{V}$ at $t^h \leq T$. 
The intensity function of an arbitrary item $v_j$ in $\mathcal{S}_{u_i}(T)$ is defined as:
\begin{equation*}
\small
\begin{split}
\lambda_{v_j}(t) = \mu_{v_j} +  \sum\nolimits_{t^h<t}\alpha_{(v_j,v^h)}\kappa(t-t^h),
\end{split}
\label{hawkes}
\end{equation*}
where $\mu_{v_j}$ (a.k.a base intensity) is a positive parameter which is independent of the interaction sequence $\mathcal{S}_{u_i}(T)$, $\alpha_{(v_j,v^h)}$ is also a positive parameter that estimates the influence between item pair $(v_j,v^h)$ and the $\kappa(t-t^h)$ is a triggering kernel function. The intensity function explicit models dynamic influence among interaction sequence. However, most existing approaches ignore to model the topological structure between a series of interaction sequences. To fill this gap, DSPP includes a novel TFE module which provides a strong structure prior to enhance model expressiveness.

\begin{figure}[t]
\begin{center}
\includegraphics[height=9cm]{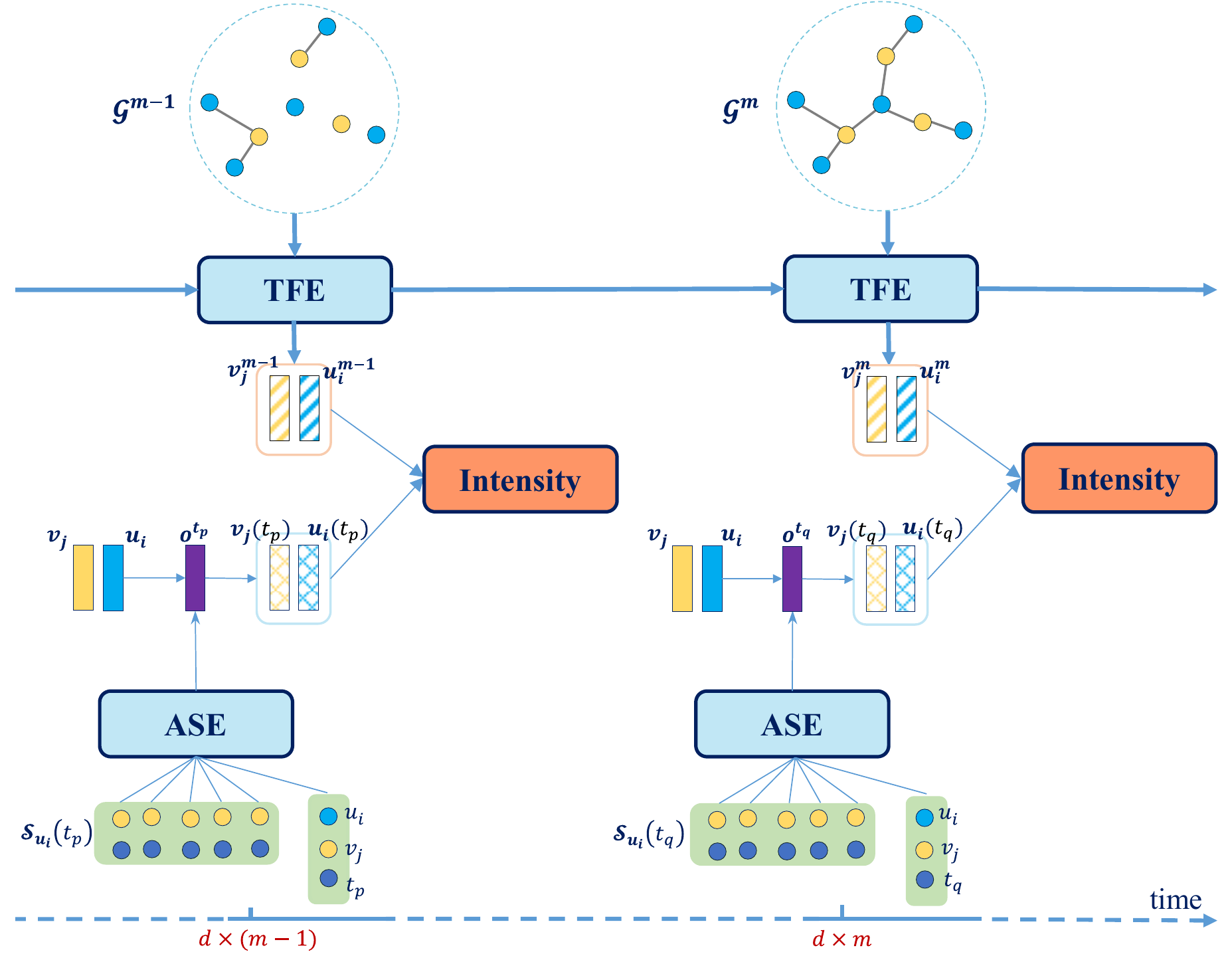}
\caption{A simple overview of DSPP. ``TFE", ``ASE", ``Intensity" mean topological fusion encoder, attentive shift encoder and intensity function, respectively. $\mathcal{G}^{m-1}$ and $\mathcal{G}^m$ are two network snapshots. Noticeably, $\mathcal{G}^{m-1} \in \mathcal{G}^{m}$.  $(u_i,v_j,t_p)$, $(u_i, v_j, t_q)$ denote two interactions, where $t_p \in [d \times (m-1), d \times m)$ and $t_q \in [d \times m, d \times (m+1))$. $\mathcal{S}_{u_i}(t_p)$ and $\mathcal{S}_{u_i}(t_q)$ are two interaction sequences.} 
\label{model}
\end{center}
\end{figure}

\section{Proposed Model}

\subsection{Overview}
For a temporal interaction network $\mathcal{G}(T) = (\mathcal{U},\mathcal{V},\mathcal{E})$, the adjacency matrix would change over time, because the emerge of a new interaction would introduce a new edge in the temporal interaction network, causing the huge occupation of memory. To sidestep this problem, we exploit an ordered snapshot sequence $\{\mathcal{G}^m\}_{m=0}^{M-1}$ with the same time interval $d = \frac{T}{M}$ to simulate the temporal interaction network $\mathcal{G}(T)$. Each snapshot $\mathcal{G}^m$ equals to $\mathcal{G}(d\times m)$, and $M$ is a hype-parameter to control the number of snapshot. 
\par
Fig.\ref{model} shows the overview of our model. In DSPP, the TFE module aims to learn \textbf{steady embeddings} which represent the stable intentions of users and items in the corresponding time period $[d \times m, d \times (m + 1))$ from $\mathcal{G}^m$. Here we denote the steady embeddings of an arbitrary user $u_i$ and an item $v_j$ as $\bm{u}^m_i$ and $\bm{v}^m_j$, respectively. 
The second module ASE aims to learn \textbf{dynamic embeddings} of users and items for describing their dynamic intentions at timestamp $t$. 
\subsection{Embedding Layer}
The embedding layer is used to initialize node embeddings (all users and items) and time embedding. It consists of two parts: the node embedding layer and the time embedding layer.
\subsubsection{Node Embedding Layer}
The node embedding layer aims to embed users and items into a low-dimensional vector space. Formally, given a user $u_i$ or an item $v_j$, we can obtain its $D$-dimensional representation ($\bm{u}_{i} \in \mathbb{R}^D$ or $\bm{v}_{j} \in \mathbb{R}^D$) from an initialization embedding matrix with a simple lookup operation, where $D$ denotes the dimension number of embeddings.

\subsubsection{Time Embedding Layer}
Position embedding~\cite{sasrec,dysat} is widely used to recognize the ordered information of sequences. However, the continuous time information of interaction sequence cannot be well reflected by the discrete position embedding. Therefore, we design a time embedding layer that encodes the discrete ordered information and continuous time information simultaneously. Concretely, given a future timestamp $t^+ \geq t$, user $u_i$ and the $h$-th interaction $(v^h,t^h)$ of its interaction sequence $\mathcal{S}_{u_i}(t)$ (detailed in Sec. \ref{TPP}), the $h$-th interaction time embedding $\bm{p}_{t^h}(t^+)$ of future timestamp $t^+$ can be formulated as follows:
\begin{equation*}
\small
\begin{split}
\left[\bm{p}_{t^h}(t^+) \right]_j =\left\{
\begin{aligned}
&\cos(\omega_j (t^+ - t^h) + h / 10000^{\frac{j-1}{D}}), \rm{if\;}\textsl{j}\rm{\;is\;odd},  \\
&\sin(\omega_j (t^+ - t^h) + h / 10000^{\frac{j}{D}}),   \rm{if\;}\textsl{j}\rm{\;is\;even}, \\
\end{aligned}
\right.
\end{split}
\label{time_encoding}
\end{equation*}
where $[\bm{p}_{t^h}(t^+)]_j$ is the $j$-th element of the given vector $\bm{p}_{t^h}(t^+)$, and $\omega_j$ is a parameter to scale the time interval $t^+ - t^h$ in the $j$-th dimension.

\subsection{Topological Fusion Encoder}
In this section, we propose a novel topological fusion encoder (TFE) to learn the topological structure. The existing graph encoders~\cite{gcrn,tgat} learn node embeddings by aggregating the features of neighboring nodes directly. However, in temporal interaction networks, it would easily lead to improper feature aggregations due to the fact that the user neighbors are items. In this paper, we introduce a topological aggregation layer (TAL) into TFE to alleviate this issue.

\subsubsection{Topological Aggregation Layer}
\begin{figure}[tbp]
\begin{center}
\includegraphics[height=4.5cm]{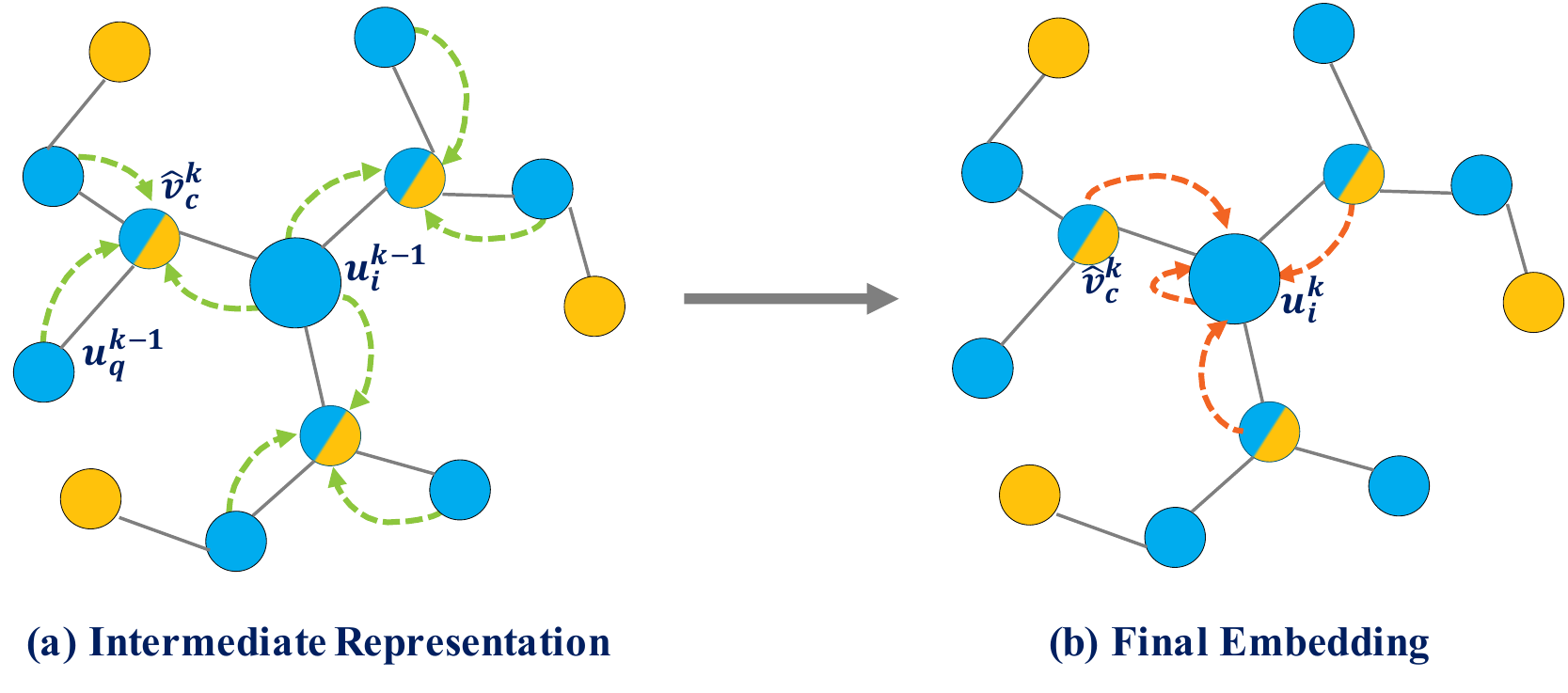}
\caption{Illustration of topological aggregation layer (TAL). Blue and yellow color nodes denote users and items, respectively. Nodes with two colors denote the intermediate representations, and gray lines denote that users have interacted with items. The sub-graphs (a) and (b) show the learning procedures of $\bm{u}^{k-1}_{i}$ in $k$-th TAL. The green dotted lines (Eq.(\ref{graph_conv_1})) and orange dotted lines (Eq.(\ref{graph_attention_1})) describe how to derive the embedding $\bm{u}_i^{k}$ by considering the topological structure of temporal interaction network.} 
\label{gnn}
\end{center}
\end{figure}

Different with homogeneous graphs, the distance between a user (item) and other users (items) is always an even number in temporal interaction network, e.g. 2, 4 and 6. This fact indicates that our encoder should aggregate homogeneous information through the even-number-hop. Based on this topological structure, we design a novel topological aggregation layer (TAL) as shown in Fig.\ref{gnn}. Concretely, given an interaction $(u_i,v_j,t)$, we firstly compute its corresponding snapshot identifier $m = \lfloor \frac {t}{d}\rfloor$, where $\lfloor \cdot \rfloor$ denotes the floor function and $d$ is the time interval of ordered snapshot sequence. Then, to generate the user representation $\bm{u}_i^k$ in the $k$-th TAL, we calculate its intermediate representation $\widehat{\bm{v}}^{k}_{c}$ as follows:
\begin{equation}
\small
\begin{split}
\widehat{\bm{v}}^{k}_{c} = \delta \Big( \widehat{W}^{k}_{u} \ {\rm MEAN} \big( \{ \bm{u}^{k-1}_{q} : u_q \in \mathcal{N}_m(v_c) \} \big) \Big), {\rm where} \ v_c \in \mathcal{N}_m(u_i),
\label{graph_conv_1}
\end{split}
\end{equation}
where $\delta$ is the ReLU activity function, $\widehat{W}^{k}_{u}$ is a parameter matrix, and $\mathcal{N}_m(v_c)$ denotes the set of 1-hop neighbors (user-type) of item $v_c$ in $\mathcal{G}^m$. Therefore, $\widehat{\bm{v}}^{k}_{c}$ can be consider as a user-type representation since it only aggregates the user-type features. After obtaining the intermediate representation $\widehat{\bm{v}}^{k}_{c}$, we leverage the attention mechanism~\cite{attention} to learn different weights among the neighboring intermediate representations for user $u_i$. The final embedding $\bm{u}^{k}_{i}$ can be formulated as:
\begin{equation}
\small
\begin{split}
e_{c} = \delta \big( (\overline{W}^{k}_{u}& \bm{u}_i^{k-1})^\top \widehat{\bm{v}}^{k}_{c} \big), {\rm where} \ v_c \in \mathcal{N}_m(u_i),\\
\alpha_{c} &=  \frac{\exp{(e_{c})}}{\sum_{v_q \in \mathcal{N}_m(u_i)}\exp{(e_{q})}},\\
\overline{\bm{u}}^{k}_{i} &= \delta \big( \sum\nolimits_{v_c \in \mathcal{N}_m(u_i)} \alpha_{c} \widehat{\bm{v}}^{k}_{c}  \big),\\
&\bm{u}^{k}_{i} = W^{k}_{u} \ \big[\overline{\bm{u}}^{k}_{i} \big| \bm{u}^{k-1}_{i} \big],
\end{split}
\label{graph_attention_1}
\end{equation}
where the $\overline{W}^{k}_{u}$ and $W^{k}_{u}$ are parameter matrices,  $(\cdot)^\top$ is the transpose operation and $[\cdot | \cdot]$ is concatenation operation. Analogously, we can employ the same learning procedure to update $\bm{v}_j^{k-1}$. In TFE, we stack $K$ TALs and denote the final outputs of $\bm{u}_i^{K}$ and $\bm{v}_j^{K}$ as our steady embeddings $\bm{u}^m_i$ and $\bm{v}^m_j$ for user $u_i$ and item $v_j$ in $\mathcal{G}^m$, respectively. 

\subsubsection{Temporal Fusion Layer}
The proposed TAL can effectively deal with a single network snapshot, but it cannot capture the structural variations across the ordered snapshot sequence $\{\mathcal{G}^0, \mathcal{G}^1, ..., \mathcal{G}^{M-1}\}$. To mitigate this problem, after obtaining user and item embeddings (e.g. $\bm{u}^m_i$ and $\bm{v}^m_j$) for each discrete snapshot $\mathcal{G}^m$, we introduce a temporal fusion layer to encode these dynamical changes in the ordered snapshot sequence:
\begin{equation}
\small
\begin{split}
\bm{u}^{m}_{i} = f_u(\bm{u}^{m-1}_i, \bm{u}^m_i),\quad\bm{v}^{m}_{j} = f_v(\bm{v}^{m-1}_j, \bm{v}^m_j),
\label{merge_1}
\end{split}
\end{equation}
where $f_u$ and $f_v$ are temporal modeling functions. There are many alternative methods that can be used for concrete implementations. In our model, we choose two separate GRUs~\cite{gru} to model $f_u$ and $f_v$, respectively.

\subsection{Attentive Shift Encoder}
In this section, we develop an attentive shift encoder (ASE) for temporal interaction networks for capturing the long-range dependency structure. Previous works employ different RNN variants which tend to forget the long history information, leading to the problem of long-range dependency structure missing. In contrast, our ASE module can explicitly learn the dependencies between each historical interaction and the new-coming interaction via the attention mechanism.

\subsubsection{Attentive Interaction Layer}
Considering a new-coming interaction $(u_i, v_j, t)$ that user $u_i$ has an interaction with item $v_j$ at the timestamp $t$, we can use it to generate the dynamic embeddings of users and items and compute the correlation among historical interactions in $\mathcal{S}_{u_i}(t)$. The concrete implementation is given as:
\begin{equation}
\small
\begin{split}
	e_{h} = \big(W_{Q}[\bm{u}_i|(\bm{v}_j + \bm{p}_{t^{|\mathcal{H}_{u_i}|}}(t&))]\big)^\top W_{K}[\bm{u}_i|(\bm{v}^h + \bm{p}_{t^h}(t))], {\rm where} \ (v^h, t^h) \in \mathcal{S}_{u_i}(t) \\
	\alpha_{h} &= \frac{\exp{(e_h)}}{\sum_{(v^c, t^c) \in \mathcal{S}_{u_i}(t)} \exp{(e_c)}}, \\
	\bm{o}^t = &\delta \big( \sum\nolimits_{(v^h, t^h) \in \mathcal{S}_{u_i}(t)} \alpha_{h} W_{V} \bm{v}^h \big),
\end{split}
\label{attention_score}
\end{equation}
where $|\mathcal{H}_{u_i}|$ is the number of interaction sequence $\mathcal{S}_{u_i}(t)$, and $\bm{o}^t$ is the new-coming interaction feature. $W_Q, W_K$ and $W_V$ are the query, key and value parameter matrices, respectively. Afterwards, we generate the embeddings of user $u_i$ and item $v_j$ at timestamp $t$ via the following operations:
\begin{equation}
\small
\begin{split}
\bm{u}_{i}(t) = g_u(\bm{u}_i, \bm{o}^t),\quad\bm{v}_{j}(t) = g_v(\bm{v}_j, \bm{o}^t),
\label{merge_2}
\end{split}
\end{equation}
where $g_u$ and $g_v$ are embedding generation functions. In our model, we also use two separate GRUs for their implementations.

\subsubsection{Temporal Shift Layer}
Intuitively, the embeddings of user and item should be changed over time. For example, electronic products will gradually reduce their prices over time, and users may have different intentions when they returned to the E-commerce platform again. Hence, maintaining the same embeddings in a period cannot reflect the reality for the future prediction~\cite{latent}. In our work, we devise a temporal shift layer to achieve dynamic embeddings over time. Specifically, after obtaining the embeddings of user $u_i$ and item $v_j$ at timestamp $t$, i.e., $\bm{u}_{i}(t)$ and $\bm{v}_{j}(t)$ in Eq.(\ref{merge_2}), their dynamic embeddings at future timestamp $t^+ \geq t$ can be calculated as follows:
\begin{equation}
\small
\begin{split}
	\bm{u}_{i}(t^+) =\left(1 + \Delta \bm{w}_{u_i}\right) * \bm{u}_{i}(t), \quad
	\bm{v}_{j}(t^+) =\left(1 + \Delta \bm{w}_{v_j}\right) * \bm{v}_{j}(t), 
\end{split}
\end{equation}
where $\Delta\!= t^+\!-t$ is the shift time interval, $*$ is the element-wise product, $\bm{w}_{u_i}$ and $\bm{w}_{v_j}$ are corresponding learnable shift vectors of user $u_i$ and item $v_j$, respectively. We assume that the user or item embedding can shift in continuous space with its own trajectory, so each user or item has a specific shift vector.

\subsection{Model Training}
To explicitly capture dynamic influence between users and items, we devise a novel intensity function which is generated via the steady embeddings and dynamic embeddings. 

\subsubsection{Intensity Function}
We model all possible interactions for all users with items via a multi-dimensional intensity function, where each user-item pair holds one dimension. Formally, based on the learned user and item embeddings, the intensity function of user-item pair $(u_i,v_j)$ is defined as follows:
\begin{equation}
\small
\begin{split}
	\lambda_{(u_i, v_j)}(t) &=\sigma \big(\underbrace{(\bm{u}_{i}^{\lfloor \frac{t}{d}\rfloor})^\top \ \bm{v}_{j}^{\lfloor \frac{t}{d} \rfloor}}_{\text{base intensity} \atop \text{(TFE)}} + \underbrace{(\bm{u}_{i}(t))^\top \ \bm{v}_{j}(t)}_{\text {dynamic change} \atop \text{(ASE)}}\big),
\end{split}
\label{intensity}
\end{equation}
where $\lfloor \frac{t}{d} \rfloor$ denotes the corresponding network snapshot identifier and $\sigma$ is the softplus function for ensuring that the intensity function is positive and smooth. Our intensity function is similar with MHP (detailed in Sec. \ref{TPP}): 1) The former term $(\bm{u}_{i}^{\lfloor \frac{t}{d}\rfloor})^\top \ \bm{v}_{j}^{\lfloor \frac{t}{d} \rfloor}$ is provided by TFE, which uses the topological structure of temporal interaction network as a strong prior to generate the base intensity. 2) The latter term $(\bm{u}_{i}(t))^\top \ \bm{v}_{j}(t)$ is obtained by ASE, which describes the dynamic changes for this user-item pair.

\subsubsection{Objective Function}
Based on the proposed intensity function, we can train our model by maximizing the log-likelihood of these happened interactions during time window $[0, T)$:
\begin{equation}
\small
\begin{split}
\mathcal{L}=\sum_{(u_i,v_j,t) \in \mathcal{G}(T)}& \!\!\!\!\!\!\!\!\log (\lambda_{(u_{i}, v_{j})}(t))-\!\!\!\!\!\!\!\!\!\!\!\!\!\!\underbrace{\int_{0}^{T} \lambda (t) \mathrm{d} t,}_{\text {non-happened interactions}}
\end{split}
\label{loss}
\end{equation}
\begin{equation}
\small
\begin{split}
\lambda(t) &= \sum\nolimits_{u_i \in \mathcal{U}} \sum\nolimits_{v_j \in \mathcal{V}}\lambda_{(u_i,v_j)}(t).
\end{split}
\label{original}
\end{equation}
Maximizing the likelihood function $\mathcal{L}$ can be interpreted intuitively in the following way: 1) The first term ensures that all happened interactions probabilities are maximized. 2) The second term penalizes the sum of the log-probabilities of infinite non-happened interactions, because the probability of no interaction happens during $[t, t+\mathrm{d}t)$ is $1 - \lambda(t)\mathrm{d}t$, and its log form is $-\lambda(t)\mathrm{d}t$~\cite{nhp}. 

\begin{algorithm}[t]
\caption{The training procedure of DSPP.}
\label{dspp}
\begin{algorithmic}[1]
\REQUIRE The training temporal interaction network $\mathcal{G}(T_{tr})$, the ordered snapshot sequence $\{\mathcal{G}^0,\mathcal{G}^1,...,\mathcal{G}^{M-1}\}$, the time interval $d$, the user set $\mathcal{U}$, the item set $\mathcal{V}$, sampling number $N$.
\STATE Initialize model parameters.

\WHILE {not convergence} 
\STATE Enumerate a batch of consecutive interactions from $\mathcal{G}(T_{tr})$ as $B$.

\STATE $\nabla \leftarrow 0$ \quad $\backslash\backslash$ Happened interactions.
\STATE $\Lambda \leftarrow 0$\quad\enspace $\backslash\backslash$ Non-happened interactions.

\FOR {each interaction $(u_i,v_j,t) \in B$}
\STATE $m$ $\leftarrow$ $\lfloor \frac{t}{d} \rfloor$ $\backslash\backslash$ Calculate the snapshot identifier.
\STATE Calculate steady embedding $\bm{u}^{m}_{i}$ and $\bm{v}^{m}_{j}$ via TFE based on $\mathcal{G}^{m}$.
\STATE Calculate dynamic embedding $\bm{u}_{i}(t)$ and $\bm{v}_{j}(t)$ by ASE.
\STATE $\nabla \leftarrow \nabla + \log(\lambda_{(u_{i}, v_{j})}(t))$
\IF {$u_i$ has next interaction at future timestamp $t^+$ with item $v_c$}
\STATE $\nabla \leftarrow \nabla + \log(\lambda_{(u_{i}, v_{c})}(t^+))$
\STATE Uniformly sample a timestamp set $t^s=\{t^s_k\}_{k=1}^{N} \leftarrow \mathrm{Uniform}(t,t^+,N)$.
\STATE Sample the negative item set $\Upsilon$ $\backslash\backslash$ Negative sampling.
\FOR{$k \in \{2,...,N\}$}
\STATE $\Lambda \leftarrow \Lambda + (t^s_k - t^s_{k-1})\lambda(t^s_k)$ $\backslash\backslash$ Monte Carlo estimation.
\ENDFOR
\ENDIF

\ENDFOR
\STATE $\mathcal{L}$ $\leftarrow$ $\nabla-\Lambda$
\STATE Update the model parameters by Adam optimizer.
\ENDWHILE
\end{algorithmic}
\end{algorithm}

\subsubsection{Prediction Tasks}
Beneficial from TPP framework, DSPP can naturally tackle the following two tasks:
\begin{itemize}
\item Item prediction: Given user $u_i$ and a future time $t^+$, \textit{what is the item that this user will interact at time $t^+$?} To answer this question, we rank all items and recommend the one that has the maximum intensity:
\begin{equation}
\small
\begin{split}
\mathrm{argmax}_{v_j} \frac{\lambda_{(u_i,v_j)}(t^+)} {\sum_{v_c \in \mathcal{V}} \lambda_{(u_i,v_c)}(t^+)},
\end{split}
\label{item_expect}
\end{equation}
\item Time prediction: Given the user $u_i$, item $v_j$ and timestamp $t_n$, \textit{how long will this user interact with this item again after timestamp $t_n$?} To answer this question, we estimate the following time expectation:
\begin{equation}
\small
\begin{split}
\Delta=\int_{t_n}^{\infty} (t-t_n) f_{(u_i,v_j)}(t) \mathrm{d} t,
\end{split}
\label{time_expect}
\end{equation}
\end{itemize}
where $f_{(u_i,v_j)}(t) = S_{(u_i,v_j)}(t)\lambda_{(u_i,v_j)}(t)$ is the conditional density (details in Section \ref{TPP}) and $\Delta$ is the expectation interaction time.

\subsection{Model Analysis}

\subsubsection{Differences with Sequential Recommendation} 
Sequential recommendation methods~\cite{gru4rec,rum,sasrec,xu2019recurrent} also focus on modeling sequential user preferences. Compared with them, DSPP has the following fundamental differences: 
\begin{itemize}
    \item In the task level, DSPP concentrates on modeling the dynamic evolution of users and items in continuous time. DSPP can not only predict the next item, but also explicitly estimates the time expectation of a given user-item interaction. In contrast, sequential recommendation aims to model interaction sequences in the discrete manner. Thus, most of them ignore the timestamp information and cannot model the time distribution.
    \item In the model level, DSPP simultaneously captures the topological structure and the long-range dependency structure via our TFE and ASE modules, but sequential recommendation methods usually ignore the topology information in temporal interaction networks. 
\end{itemize}

\subsubsection{Time Complexity}
To accelerate the training process of DSPP, we adopt $t$-$batch$ algorithm~\cite{jodie} to organize data for paralleling training. Moreover, we apply Monte Carlo Algorithm~\cite{survival} with the negative sampling trick~\cite{coeve} to estimate our objective function Eq.(\ref{loss}). Hence, the main operations of DSPP fall into the proposed TFE and ASE modules. The computational complexity of TFE is $\mathcal{O}(K|\mathcal{E}|D)$, and the ASE is $\mathcal{O}(HBD)$, where $K$ is the number of TAL, $B$ is the batch size, and $H$ is a hype-parameter to control the maximum length of historical interactions. In general, our model keeps an efficient training speed. Empirically, in the same running environment, JODIE~\cite{jodie}, DGCF~\cite{dgcf} and DSPP would cost about 5.1 mins, 17.7 mins, and 8.15 mins per epoch on the Reddit dataset, respectively. The pseudo code of the training procedure is shown in Algorithm \ref{dspp}.


 \begin{table}[tbp]
\centering
\caption{Statistics of three datasets.}
\setlength{\tabcolsep}{15pt}
\resizebox{12cm}{!}{
\begin{tabular}{|c|c|c|c|c|}
\hline
Datasets  &  $|\mathcal{U}|$  &$|\mathcal{V}|$ &Interactions & Action Repetition \\ \hline
Reddit  &10,000  &1,000 &672,447 &79\%    \\ \hline
Wikipedia &8,227  &1,000  &157,474 &61\%     \\ \hline
Last.FM &1,000  &1,000  &1,293,103  &8.6\%      \\ \hline
\end{tabular}
}
\label{dataset}
\end{table}
\section{Experiments}
\subsection{Datasets}
To make a fair comparison, we evaluate DSPP on three pre-processed benchmark datasets~\cite{jodie}, i.e., Reddit\footnote{\url{http://snap.stanford.edu/jodie/reddit.csv}}, Wikipedia\footnote{\url{http://snap.stanford.edu/jodie/wikipedia.csv}} and Last.FM\footnote{\url{http://snap.stanford.edu/jodie/lastfm.csv}}. The concrete statistics of users, items, interactions and action repetition are listed in Table \ref{dataset}. Noticeably, these datasets are largely different in terms of action repetition rate, which can verify whether DSPP is able to capture the dynamic influence in various action repetition scenarios accurately.




\subsection{Experiment Setting}
\textbf{Data Preprocessing}: As used in JODIE~\cite{jodie} and DGCF~\cite{dgcf}, for each dataset, we first sort all interactions by chronological order. Then, we use the first 80\% interactions to train, the next 10\% interactions to valid, and the remaining 10\% interactions for the test. In contrast with JODIE and DGCF, we generate a snapshot sequence $\{\mathcal{G}^m\}_{m=0}^{M-1}$. In our setting, the validation snapshots cover the training data, and the test snapshots also contain all training and validation data.
\\
\textbf{Evaluation Metrics}: To evaluate our model performance, for each interaction, we first generate corresponding user and item steady and dynamic embeddings. Then, we rank all items by Eq.(\ref{item_expect}) and predict the future time by Eq.(\ref{time_expect}). Afterward, we evaluate the item prediction task with the following metrics: Mean Reciprocal Rank (MRR) and Recall@10. higher values for both metrics are better. For the time prediction task, we use Root Mean Square Error (RMSE) to measure model performance, and a lower value for RMSE is preferred.
\\
\textbf{Baselines}: We compare DSPP with the following baselines.
\begin{itemize}
	\item Random walk model: CTDNE~\cite{ctdne}.
	\item Recurrent network models: LSTM~\cite{lstm}, RRN~\cite{rrn}, LatentCross~\cite{latent}, Time-LSTM~\cite{timelstm}, JODIE~\cite{jodie} and DGCF~\cite{dgcf}.
	\item Temporal point process model: DeepCoevolve~\cite{coeve}.
\end{itemize}
\textbf{Implementation Details}: In our experiments, we use the official implementations\footnote{\url{https://hanjun-dai.github.io/supp/torch_coevolve.tar.gz}} of DeepCoevolve. Except from it, we directly report the experimental results in the original papers~\cite{jodie,dgcf}. DSPP follows the same hyper-parameter setting with baselines: the embedding dimension $D$ is fixed as 128, the batch size $B$ is fixed as 128, the learning rate is fixed as 0.001, the model weight decay is fixed as 0.00001, the sampling number for Monte Carlo estimate is fixed as 64, the number of negative sampling is fixed as 10, the number of TAL $K$ is fixed as $2$, the Attention is stacked 8 layers, the GRUs are 1 layer, the number of snapshots $M$ is selected from $\{128, 256, 512, 1024\}$ and the maximum length of interaction sequence $H$ is chosen from $\{20, 40, 60, 80\}$. The Adam~\cite{adam} optimizer is used to update all model parameters.
\begin{table}[tbp]
\centering
\caption{Performance (\%) comparison of item prediction.}
\setlength{\tabcolsep}{20pt}
\resizebox{12cm}{!}{
  \begin{tabular}{cp{10pt}p{10pt}p{10pt}p{10pt}p{10pt}p{10pt}}
    \toprule
    \multirow{2}{*}{Model} & \multicolumn{2}{c}{Last.FM} & \multicolumn{2}{c}{Wikipedia}  & \multicolumn{2}{c}{Reddit}\\ \cmidrule(lr){2-3}\cmidrule(l){4-5}\cmidrule(l){6-7}
    &Recall@10 &MRR &Recall@10 &MRR &Recall@10 &MRR \\  \midrule
    CTDNE & 1.0 & 1.0 &5.6  &3.5   &25.7  &16.5\\
    LSTM &12.7  &8.1 &45.9  &33.2  &57.3  &36.7  \\
    Time-LSTM &14.6  &8.8  &35.3  &25.1  &60.1  &39.8  \\
    RRN &19.9 &9.3 &62.8 &53.0 &75.1 &60.5 \\
   	LatentCross &22.7  &14.8  &48.1 &42.4    &58.8 &42.1 \\
    DeepCoevolve &33.6  &21.3  &60.6 &48.5 &78.7  &65.4   \\
    JODIE  &38.7  &23.9  &82.1  &74.6  &85.1  &72.4  \\
    DGCF  &45.6  &32.1  &85.2  &78.6  &85.6  &72.6  \\
	\midrule
    DSPP  & \textbf{47.1} & \textbf{34.3} & \textbf{90.5} & \textbf{82.1} & \textbf{86.7} & \textbf{74.5}  \\ \bottomrule
    \end{tabular}
}
\begin{center}
\end{center}
\label{nextitem}
\end{table}

\subsection{Item Prediction}
For item prediction, Table \ref{nextitem} shows the comparison results on the three datasets according to Recall@10 and MRR. From the experimental results, we have the following observations: 
\begin{itemize}
\item DSPP consistently yields the best performances on all datasets for both metrics. Compared with state-of-the-art baselines, the most obvious effects are that DSPP achieves the 6.8\% improvement in terms of MRR on Last.FM, the 6.2\% improvement in terms of Recall@10 on Wikipedia and the 2.6\% improvement in terms of MRR on Reddit. It reveals that incorporating the topological structure and long-range dependency structure can bring good robustness in different action repetition scenarios.
\item DSPP outperforms DeepCoevolve significantly. This phenomenon demonstrates that DSPP has a more powerful intensity function that can better capture the dynamic influence between users and items.	
\item DSPP and DGCF are superior to other baselines on Last.FM, which indicates that it is critical to model the topological structure information for learning temporal interaction networks.
\end{itemize}

\begin{table}[tbp]
\centering
\caption{Performance (hour$^2$) comparison of time prediction.}
\setlength{\tabcolsep}{30pt}
\resizebox{12cm}{!}{
  \begin{tabular}{cccc}
    \toprule
    Model  &Last.FM &Wikipedia &Reddit\\  \midrule
    DeepCoevolve &9.62  &10.94  &11.07\\
    DSPP  &7.78 &8.71 &9.06 \\ \bottomrule
    \end{tabular}
}
\begin{center}
\end{center}
\label{timepred}
\end{table}

\subsection{Time Prediction}
Table \ref{timepred} shows the prediction performances of DeepCoevolve and DSPP. From it, we can observe that DSPP achieves more accurately time prediction. Specifically, our model achieves the 19.1\% improvement on Last.FM, the 20.3\% improvement on Wikipedia and the 18.1\% improvement on Reddit. We suppose that the improvements owe to the following reasons: 1) DeepCoevolve uses a linear intensity function to model dynamic influence over time, which would reduce the model flexibility of intensity function. 2) DeepCoevolve remains the same user and item embeddings until it involves a new-coming interaction, so it limits model expressiveness. In contrast, our intensity function can learn the nonlinear dynamic influence, since the ASE module can provide time-aware dynamic embeddings.

\begin{table}[tbp]
 \centering
 \caption{Performance (\%) comparison of different model variants.}
 \setlength{\tabcolsep}{35pt}
 \resizebox{12cm}{!}{
 \begin{tabular}{lcc}  
  \toprule
  Model  & Recall@10 & MRR \\
  \midrule
  DSPP        & 47.1 & 34.3 \\
  \enspace  \enspace  Remove TFE & 40.2 & 25.3 \\
  \enspace  \enspace  Replace TAL with GCN  & 44.5 & 32.4 \\
  \enspace  \enspace  Replace TAL with GAT  & 45.2 & 32.7 \\
  \bottomrule
 \end{tabular}
 }
\label{variant}
\end{table}

\subsection{Discussion of Model Variants}
To investigate the effectiveness of our model components, we implement several variants of DSPP and conduct the experiment on Last.FM dataset for the task of item prediction. The experimental results are reported in Table \ref{variant}. 
According to it, we can draw the following conclusions: 
\begin{itemize}
\item Remove TFE. To verify whether our proposed TFE module is useful to enhance the expressiveness of the intensity function, we first remove it and only remain the second term of Eq.(\ref{intensity}) as our intensity function. Then, we directly predict the item that is most likely to interact with each user via Eq.(\ref{item_expect}). As shown in the results, both metrics Recall@10 and MRR sharply drop 14.6\% and 26.2\%, respectively. It demonstrates that modeling the topological structure of temporal interaction networks can provide a powerful structural prior for enhancing the expressiveness of intensity function.
\item Remove TFE. This variant can be also viewed as a non-graph based model, since it does not exploit the topological structure, and the remaining temporal attention shift encoder only provides the long-range dependency structure to model intensity function. Compared with DeepCoevolve, this variant yields 19.6\% and 18.7\% improvements on Recall@10 and MRR respectively. This observation shows that our proposed temporal attention shift encoder can further enhance the intensity function. 
\item Replace TAL with GCN/GAT. To verify whether our proposed TAL is superior to other graph encoders for capturing the topological structure of temporal interaction networks. We replace TAL by GCN~\cite{gcn} and GAT~\cite{gat}. For the GCN variant, both Recall@10 and MRR drop 5.8\%. For GAT variant, Recall@10 and MRR drop 4.0\% and 4.6\%, respectively. So, We suppose that our proposed TAL can better capture the information of the same type entity.
\end{itemize}

\section{Conclusion}
In this paper, we present the deep structural point process for learning temporal interaction networks. Our model includes two proposed modules, i.e., topological fusion encoder and attentive shift encoder to learn the topological structure and the long-range dependency structure in temporal interaction networks, respectively. On top of that, a novel intensity function, which combines the learned steady and dynamic embeddings, is introduced to enhance the model expressiveness. Empirically, we demonstrate the superior performance of our model on various datasets for both tasks of item prediction and time prediction. 

\section*{Acknowledgement}
This work is supported in part by the Strategic Priority Research Program of Chinese Academy of Sciences (Grant No. XDC02040400) and the Youth Innovation Promotion Association of CAS (Grant No. 2021153).
%
%

\end{document}